\documentclass[pra, showpacs, twocolumn, floatfix]{revtex4}
\usepackage{graphicx}
\usepackage{amsmath, amsfonts, amssymb, bm}
\begin{document}
\title{Loading atom lasers by collectivity-enhanced optical pumping}

\author{Mihai A. \surname{Macovei}}
\email{mihai.macovei@mpi-hd.mpg.de}

\author{J\"{o}rg \surname{Evers}}
\email{joerg.evers@mpi-hd.mpg.de}

\affiliation{Max-Planck-Institut f\"{u}r Kernphysik, Saupfercheckweg
1, D-69117 Heidelberg, Germany}
\date{\today}

\begin{abstract}
The effect of collectivity on the loading of an atom laser via optical
pumping is discussed. In our model, atoms in a beam are laser-excited 
and subsequently spontaneously decay into a trapping state. We consider the
case of sufficiently high particle density in the beam such that 
the spontaneous emission is modified by the particle interaction.
We show that the collective effects lead to a better population of
the trapping state over a wide range of system parameters, and that
the second order correlation function of the atoms can be controlled
by the applied laser field.
\end{abstract}
\maketitle
\section{Introduction}

Motivated by the many interesting features and applications of both Bose-Einstein condensates (BEC) and of optical lasers, significant effort is devoted towards the creation of a continuous atom laser. An atom laser is an intense coherent matter wave, and is typically extracted from a Bose-Einstein condensate~\cite{bec1,bec2}. Proposed applications include precision measurements and fundamental tests of quantum mechanics~\cite{appl}. So far, only pulsed atom lasers could be realized experimentally~\cite{pal,scoh,bloch,qal,hsl,pbr,guide}.

Recently, the second-order correlation function of an atom laser could be measured in a Hanbury Brown-Twiss type experiment~\cite{scoh}. In~\cite{qal}, a quasi continuous atom laser was demonstrated. Similar, the Heisenberg limit could be approached in an atom laser~\cite{hsl}, and high peak brightness was achieved in~\cite{pbr}. The first experiment on a guided quasicontinuous atom laser was performed in \cite{guide}. 
Experimental realization of a multi-beam atom laser was reported in~\cite{mat}, while interference of an array of atom lasers was observed in~\cite{int}. The propagation of atom laser beams is discussed~\cite{pral}, and the steady-state quantum statistics of a non-Markovian atom laser was investigated in~\cite{ssnm}. Atom laser coherence and its control via feedback was analyzed in~\cite{atc}. Also paired-atom laser beams created via four-wave mixing were discussed~\cite{pat}. A scheme for creating quadrature- and intensity-squeezed atom lasers that do not require squeezed light as an input was described in~\cite{sqa}.

So far, however, no continuous atom laser could be realized, despite significant effort. For example, a continuous source of BEC atoms was obtained in \cite{cws}. Promising mechanisms for providing a pumping mechanism consistent with a continuous atom laser have recently been demonstrated~\cite{patl}. Loading a continuous-wave atom laser by optical pumping techniques was shown in \cite{thrl} while the continuous pumping of atoms into a BEC via spontaneous emission from a thermal reservoir of atoms was investigated in \cite{san}. Continuous loading of a non-dissipative 
atom trap was studied in \cite{cwl}. Stability of continuously pumped atom lasers was discussed as well \cite{st}. 
In~\cite{cwal}, an atom laser that is simultaneously pumped and output-coupled to a free beam was achieved. 

\begin{figure}[b]
\includegraphics[width=6.5cm]{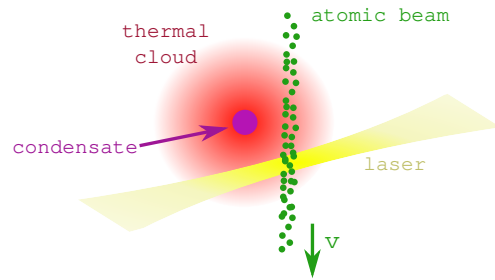}
\caption{\label{fig-1}(Color online) The scheme for obtaining a steady-state BEC and a cw atom laser. }
\end{figure}

The optical pumping techniques in~\cite{thrl} aim at loading atoms into a magnetic trap without heating the system or destroying the condensate. For this, a laser field is used to pump atoms initially in a ground state to an excited state, which subsequently decays into a trapping state, see Fig.~\ref{fig-1}. The authors showed that photon reabsorptions, which are a major limitations to such schemes as they can remove atoms from the cloud or lead to a heating, could be minimized by adjusting the system geometry and by inducing spontaneous emission frequencies which do not coincide with the resonance frequencies for reabsorption. Overall, the figure of merit is the final population in the trapping state. 

It is well known that collective interactions between closely spaced particles can lead to a significant modification of spontaneous emission processes. The quantum dynamics of a collective system can be $N$ times faster than for a single particle, and the intensity of the emitted electromagnetic field scales as $N^{2}$ in multiparticle samples, where $N$ is the number of atoms~\cite{dicke}. It has been shown that the collective dynamics can be controlled~\cite{mekPRA}. Interestingly, recently a clear $N^{2}$ dependence of the fluorescence light emitted by inverted three-level $\Lambda-$ type indium atoms could be observed already at rather low densities~\cite{exsf}. Also the superradiant emission of a driven thin solid sample in an optical resonator was observed~\cite{ssm}. 

Motivated by this, here, we study the effect of collectivity on the loading of an atom laser via optical pumping.
We base our analysis on the model presented in~\cite{thrl}, but assume that the particle density in the beam is sufficiently high such that collective interactions become relevant. We discuss the influence of particle density, laser parameters, reabsorption, and the ratio of the natural spontaneous emission rates in the studied atoms on the final population in the trapping state. We find that collective effects lead to a better population of the trapping state over a wide range of system parameters, starting already at the onset of collectivity. Finally, we study the  second order correlation function of the atoms, and show that it can be controlled, e.g.,  via the detuning of the applied laser field.

\section{The Model}
\begin{figure}[t]
\includegraphics[width=4.5cm]{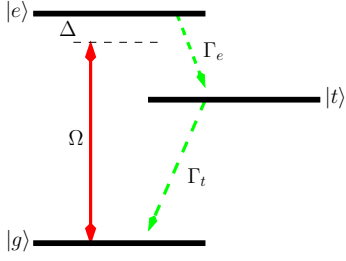}
\caption{\label{fig-2}(Color online)  Level scheme. The laser couples the 
states $|g \rangle$ and $|e \rangle$ with Rabi frequency 
$\Omega$. Spontaneous emission of photons occurs with rates 
$\Gamma_{e}$ and $\Gamma_{t}$, respectively.}
\end{figure}

We analyze a scheme for loading a thermal cloud into a magnetic trap by optically pumping atoms from an external cold atomic-beam source (see Fig.~\ref{fig-1}). In this model, atoms are injected into the trap in the state $|g\rangle$ (see Fig.~\ref{fig-2}). A laser field excites the particles when they enter the spatial region containing the thermal cloud into state $|e\rangle$. From the excited state $|e\rangle$ the atoms collectively emit photons and ideally end up in the trap state $|t\rangle$ (see Fig.~\ref{fig-1} and Fig.~\ref{fig-2}). This setup allows for evaporative cooling in the steady state~\cite{ssbc,thrl}. The atoms in the trapping state may escape the trap by absorbing photons or due to collisions. 

We model the dynamics via the  master equation
\begin{align}
\dot \rho &+ \frac{i}{\hbar}[H_{0},\rho]= -\Gamma_{e}(1+\bar n_{e})
[S_{et},S_{te}\rho]-\Gamma_{e}\bar n_{e}[S_{te},S_{et}\rho] \nonumber \\
&-\Gamma_{t}(1+\bar n_{t})[S_{tg},S_{gt}\rho]-\Gamma_{t}\bar n_{t}[S_{gt},S_{tg}\rho] +H.c.\,,
\label{me}
\end{align}
where 
\begin{align}
H_{0}=\hbar \Delta S_{ee}+\hbar \omega_{tg}S_{tt}+\hbar \Omega(S_{eg}+S_{ge})\,.
\label{ham}
\end{align}
The detuning $\Delta=\omega_{eg}-\omega_{L}$ and the overdot denotes differentiation 
with respect to time. $\omega_{\alpha \beta}=\omega_{\alpha}-\omega_{\beta}$ are transition frequencies, with $\alpha,\beta \in \{g,t,e\}$. The system of $N$ atoms is described using collective operators $S_{\alpha\beta}=\sum_{j}{S_{\alpha \beta}^{(j)}}$. Here, $S_{\alpha\beta} = \sum_{j=1}^{N}|\alpha \rangle_{j}{}_{j} \langle \beta|$, which describes populations for $\alpha=\beta$, transitions for $\alpha \not=\beta$.
The Hamiltonian Eq.~(\ref{ham}) contains free energies and transitions induced by the laser field with Rabi frequency $\Omega$. In Eq.~(\ref{me}), the terms proportional to $\Gamma_{\alpha}(1+\bar n_{\alpha})$ represent spontaneous and bath-induced transitions to the lower levels while those proportional to $\Gamma_{\alpha}\bar n_{\alpha}$ describe the bath-induced transitions to the upper states. 
We have omitted the coherent part of the dipole-dipole interaction, which is justified if the Rabi frequency $\Omega$ dominates over the dipole-dipole induced energy shifts. This in essence sets an upper bound for the particle density for a given Rabi frequency. Finally, we note that the collective atomic operators obey the commutator relation
\begin{align}
[S_{\alpha\beta },S_{\alpha^{\prime}\beta^{\prime }}] = 
\delta _{\beta\alpha^{\prime}}S_{\alpha \beta ^{\prime }} - 
\delta _{\beta^{\prime}\alpha}S_{\alpha^{\prime }\beta}\,,
\end{align}
where $\alpha,\beta \in \{g,t,e\}$.

It is convenient to work in a laser-dressed picture. For this, we represent the collective operators $S_{\alpha \beta}$ via Bose operators, i.e. $S_{\alpha \beta}=c^{\dagger}_{\alpha}c_{\beta}$ with $\{\alpha, \beta \} \in \{e,t,g\}$, and perform the dressed-state transformation 
\begin{subequations}
\begin{align}
c_{g} &= \cos{\theta}q_{-} + \sin{\theta}q_{+}\,,  \\
c_{t} &= q_{t}\,,  \\
c_{e} &= -\sin{\theta} q_{-} + \cos{\theta}q_{+}\,,
\end{align}
\end{subequations}
with 
\begin{align}
\cot{2\theta}=\frac{\Delta}{2\Omega}\,.
\label{theta}
\end{align}
Assuming again a sufficiently strong laser field, we apply the secular 
approximation, and arrive at the master equation
\begin{align}
\dot \rho &+ \frac{i}{\hbar}[\tilde H_{0},\rho] = \nonumber \\
&- \{\Gamma_{e}(1 + \bar n_{e})\sin^{2}{\theta}
+\Gamma_{t}\bar n_{t}\cos^{2}{\theta}\}[R_{-t},R_{t-}\rho] \nonumber \\
&- \{\Gamma_{e}(1 + \bar n_{e})\cos^{2}{\theta}
+\Gamma_{t}\bar n_{t}\sin^{2}{\theta}\}[R_{+t},R_{t+}\rho] \nonumber \\
&- \{\Gamma_{t}(1 + \bar n_{t})\cos^{2}{\theta}
+\Gamma_{e}\bar n_{e}\sin^{2}{\theta}\}[R_{t-},R_{-t}\rho] \nonumber \\
&- \{\Gamma_{t}(1 + \bar n_{t})\sin^{2}{\theta}
+\Gamma_{e}\bar n_{e}\cos^{2}{\theta}\}[R_{t+},R_{+t}\rho] \nonumber \\
&+ H.c. \,. \label{dme} 
\end{align}
Here,
\begin{align}
\tilde H_{0} = \hbar \omega_{tg}R_{tt} + \hbar \tilde \Omega(R_{++}-R_{--})
+\hbar \frac{\Delta}{2}(R_{++}+R_{--})\,,
\end{align}
with $\tilde \Omega=\sqrt{\Omega^{2}+(\Delta/2)^{2}}$  and $R_{\alpha \beta}
=q^{\dagger}_{\alpha}q_{\beta}$ ($\{\alpha, \beta \} \in \{+,-,t\}$).

Next, we solve the master equation Eq.~(\ref{dme}) in order to estimate the population of the trapping state. We make an ansatz for the steady-state solution in the form
\begin{eqnarray}
\rho_{s}=Z^{-1}e^{-\xi R_{++}}e^{-\zeta R_{--}} \,, \label{ss} 
\end{eqnarray}
where the normalization $Z$ is determined by the requirement $Tr(\rho_{s})=1$.
Inserting Eq.~(\ref{ss}) in Eq.~(\ref{dme}) and assuming steady state $\dot \rho =0$, one obtains:
\begin{subequations}
\label{ss-par}
\begin{align}
\xi &= \ln{\biggl [ \frac{\Gamma_{e}(1+\bar n_{e})\cos^{2}{\theta}+\Gamma_{t}\bar n_{t}\sin^{2}{\theta}}
{\Gamma_{e}\bar n_{e}\cos^{2}{\theta}+\Gamma_{t}(1+\bar n_{t})\sin^{2}{\theta}}\biggr ]} \,,  \\
\zeta &= \ln{\biggl [ \frac{\Gamma_{e}(1+\bar n_{e})\sin^{2}{\theta}+\Gamma_{t}\bar n_{t}\cos^{2}{\theta}}
{\Gamma_{e}\bar n_{e}\sin^{2}{\theta} + \Gamma_{t}(1+\bar n_{t})\cos^{2}{\theta}}\biggr ]} \,.
\end{align}
\end{subequations}
Note that leading corrections to the steady-state results obtained in the secular approximation are of the order of $N\Gamma_{\alpha}(1+\bar n_{\alpha})/\tilde \Omega$ and can be neglected in the intense field limit. 

In order to calculate the relevant expectation values, we introduce atomic states $|N,n,m\rangle$ corresponding to the su(3) algebra of the operators $R_{\alpha \beta}$ (for details see, for instance, \cite{mek}). The state $|N,n,m\rangle$ describes a system of $N$ atoms with $n$ atoms in state $|+\rangle$, $m-n$ atoms in state $|t\rangle$, and $N-m$ atoms in state $|-\rangle$. For example, we find ($k_1, k_2 \in \{ 0,1,2,\dots\}$) 
\begin{figure}[t]
\includegraphics[width=6.5cm]{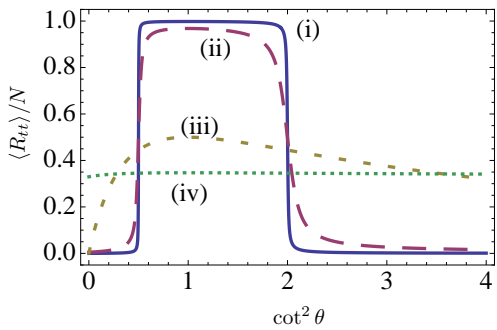} 
\caption{\label{fig-3}(Color online) Population in the trapping state as a function of the laser parameters. Curve (i) is for $N=1000$ and $\bar n_{e}=\bar n_{t}=0$, curve (ii) corresponds to $N=1000$ and $\bar n_{e}=\bar n_{t}=5$, (iii) shows $N=1$ and $\bar n_{e}=\bar n_{t}=0$, and (iv) has parameters $N=1$ and $\bar n_{e}=\bar n_{t}=5$. The ration of spontaneous emission rates is $\eta=2$. }
\end{figure}
\begin{align}
\langle R^{k_{1}}_{++}R^{k_{2}}_{--}\rangle_{s} = Z^{-1}\biggl(-\frac{\partial}{\partial \xi} \biggr)^{k_{1}}
\biggl(-\frac{\partial}{\partial \zeta} \biggr)^{k_{2}}Z \,, \label{mom}
\end{align}
where
\begin{align}
Z&=e^{-(\xi+\zeta)N}\bigl[e^{\xi(N+1)}-e^{\xi(N+2)}-e^{\zeta(N+1)}+e^{\zeta(N+2)} \nonumber \\
&+e^{(\xi+\zeta)(N+1)}(e^{\xi}-e^{\zeta})\bigr]/(e^{\xi}-1)(e^{\zeta}-1)(e^{\xi}-e^{\zeta})\,, 
\label{ssz}
\end{align}
and $\langle R_{++}\rangle + \langle R_{--}\rangle +\langle R_{tt}\rangle=N$.
In the next section, we will discuss our results based on Eqs.~(\ref{ss})-(\ref{ssz}).

\section{Results}
We start by analyzing  population of the trapping state $|t\rangle$. In Fig.~\ref{fig-3}, we plot this population as a function of $\cot^{2}{\theta}$, which depends on the laser parameters via Eq.~(\ref{theta}). We have assumed a ratio of the two incoherent decay rates $\eta=\Gamma_{e}/\Gamma_{t}=2$, and compare the single atom case $(N=1)$ to a case with strong collectivity due to a rather large number of atoms ($N=1000$). The curve (i) for $\bar{n}_e = \bar{n}_t = 0$ models the dynamics without reabsorption of the photons in the trap. This situation may occur if spontaneous photons are emitted at frequencies other than the bare transition frequencies~\cite{thrl}. We find that in this case, the population of the trapping state almost achieves the maximum value $\langle R_{tt} \rangle = N$, see the solid line in Fig.~\ref{fig-3}. If reabsorption is considered ($\bar{n}_e = \bar{n}_t = 5$), the trapping state population decreases as shown by curve (ii) in Fig.~\ref{fig-3}. The other curves (iii) and (iv) show the corresponding results for the case $N=1$ without collective interactions. 
We thus find that the collective case $N>1$ can lead to a much more efficient population of the trapping state. We will find later, however, that this result also depends on parameter $\eta$ describing the ratio of the spontaneous emission into and out of the trapping state.
\begin{figure}[t]
\includegraphics[width=6.5cm]{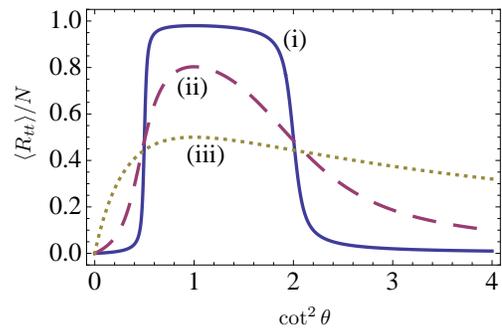}
\caption{\label{fig-4}(Color online) Population in the trapping state for different beam densities. Particle number is (i) $N=100$, (ii) $N=10$, and (iii) $N=1$.
The other parameters are $\eta=2$ and  $\bar n_{e}=\bar n_{t}=0$. }
 \end{figure}

\begin{figure}[t]
\includegraphics[width=6.5cm]{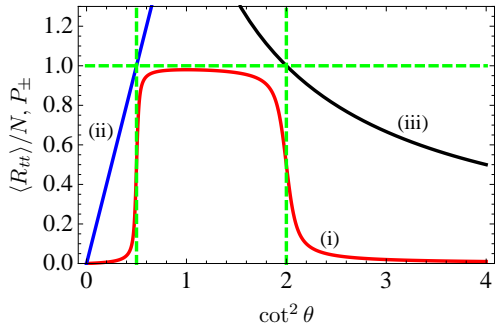}
\caption{\label{fig-interpret} (Color online) Analysis of the laser parameters leading to an efficient population of the trapping state. The dashed straight (green) lines indicate $\cot^2\theta = \eta$, $\cot^2\theta = 1/\eta$, and $\langle R_{tt}\rangle/N=1$. The red curve (i) shows the scaled population in the trapping state $\langle R_{tt}\rangle/N$ for parameters as in Fig.~\ref{fig-4}(i). The blue curve (ii) shows the ratio $P_+$, and the black curve (iii) depicts $P_-$.}
\end{figure}

In order to study the dependence of the population efficiency on the collectivity better, in Fig.~\ref{fig-4}, we plot the population of the trapping state for different numbers of atoms. It can be seen from curve (ii) that already for a rather low $N=10$, a significant increase of the trapping state population is achieved. Interestingly, from this figure, we see that an increase in the number of atoms leads to a more efficient population of the trapping state only in a finite range of $\theta$, with $0.5 \lesssim \cot^2 \theta \lesssim 2$.
This can be understood by analyzing the incoherent pumping rates into and out of the trappings state. These pumping rates $P_{ij}$ from $|i\rangle$ to $|j\rangle$
from Eq.~(\ref{dme}) follow as
\begin{subequations}
\label{pump}
\begin{align}
P_{+t} &= \Gamma_{e}(1 + \bar n_{e})\cos^{2}{\theta}
+\Gamma_{t}\bar n_{t}\sin^{2}\,,\\
P_{t+} &= \Gamma_{t}(1 + \bar n_{t})\sin^{2}{\theta}
+\Gamma_{e}\bar n_{e}\cos^{2}\,,\\
P_{-t} &=\Gamma_{e}(1 + \bar n_{e})\sin^{2}{\theta}
+\Gamma_{t}\bar n_{t}\cos^{2} \,,\\
P_{t-} &=\Gamma_{t}(1 + \bar n_{t})\cos^{2}{\theta}
+\Gamma_{e}\bar n_{e}\sin^{2} \,.
\end{align}
\end{subequations}
We define the ratios $P_+ = P_{+t}/P_{t+}$ and $P_- = P_{-t}/P_{t-}$ and show them together with the population in the trapping state in Fig.~\ref{fig-interpret}. It can be seen that the trapping state is efficiently populated if $P_+ >1$ and $P_- >1$. In this case, from Eqs.~(\ref{pump}) it follows that there is a net pumping from the laser-dressed states $|\pm\rangle$ into the trapping state. If either $P_+<1$ or $P_-<1$, a pumping channel out of the trapping state exists, and it is virtually empty. From these conditions, we can derive the range of laser parameters over which the trapping state is populated, which evaluates to
\begin{align}
\frac{1}{\eta} \leq \cot^2 \theta \leq \eta\,.
\end{align}
\begin{figure}[b]
\includegraphics[width=6.5cm]{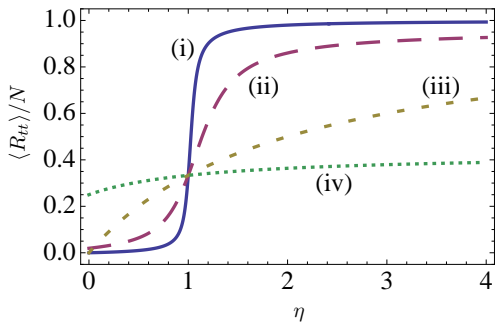}
\caption{\label{fig-ratio}(Color online) Population in the trapping state as a function of the ratio $\eta$ of the spontaneous decay rates into and out of the trapping state. The parameters are 
(i) $N=100$ and $\bar n_{e}=\bar n_{t}=0$,
(ii)  $N=100$ and $\bar n_{e}=\bar n_{t}=2$,
(iii) $N=1$ and $\bar n_{e}=\bar n_{t}=0$,
(iv)  $N=1$ and $\bar n_{e}=\bar n_{t}=2$.
The laser parameters are $\cot^{2}{\theta}=1$.}
\end{figure}

As expected, by comparing Figs.~\ref{fig-3} and \ref{fig-4} and other results not shown here, we find that increasing $N$ leads to the appearance of sharp jumps between states with almost all atoms either in or out of the trapping state. This occurs already at a moderate number of atoms. For example, increasing the number of atoms beyond $N=100$ in Fig.~\ref{fig-4} to $N=1000$ in Fig.~\ref{fig-3} induces 
only relatively small changes in the trapping population. At higher atom numbers, we also find that the influence of the repumping is suppressed, as can be seen from the different $\bar n$ in Figs.~\ref{fig-3} and \ref{fig-4} which do not lead to a strong modification as it is the case for small $N$.

Next, we analyze the trapping population as a function of the ratio $\eta$ of the spontaneous decay rates into and out of the trapping state, see Fig.~\ref{fig-ratio}. We see that the trapping state population strongly depends on $\eta$. Larger trapping state populations can be expected for $\eta > 1$, because then the decay out of the trapping state is smaller than the decay into it. Only then, increasing the sample size $N$ enhances the trapping state population.
In the opposite case $\eta <1$, only little population can be transferred into the trapping state, and increasing the sample size $N$ even can have a negative effect. Thus we conclude that efficient population of the trapping state in steady-state via collectivity requires that $\eta >1$. From Fig.~\ref{fig-ratio}, it can also be seen that for medium-sized samples, already $\eta \approx 2$ leads to almost perfect population of the trapping state if incoherent repumping is weak.
\begin{figure}[t]
\includegraphics[width=6.5cm]{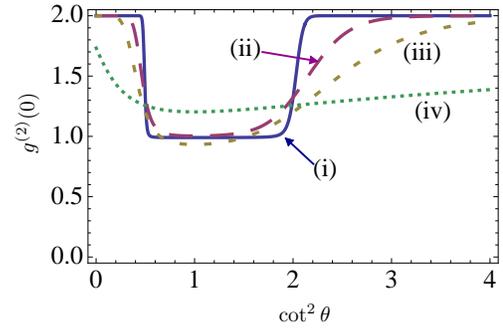}
\caption{\label{fig-g2} (Color online) Second order correlation function $g^{(2)}(0)$ of the atomic trapping state. 
(i) $N=100$ and $\bar n_{t}=\bar n_{e}=0$,
(ii) $N=100$ and $\bar n_{t}=\bar n_{e}=2$,
(iii) $N=10$ and $\bar n_{t}=\bar n_{e}=0$,
(iv) $N=10$ and $\bar n_{t}=\bar n_{e}=2$.
The spontaneous emission ratio is $\eta=2$.}
\end{figure}

We also calculated the second-order coherence function of the atoms in the trapping state $|t\rangle$, which is given by 
\begin{eqnarray}
g^{(2)}(0)=\frac{\langle q^{\dagger}_{t}q^{\dagger}_{t}q_{t}q_{t}\rangle}{\langle q^{\dagger}_{t}q_{t}\rangle^{2}} = 
\frac{\langle R_{tt}(R_{tt}-1)\rangle}{\langle R_{tt}\rangle^{2}}. \label{scf}
\end{eqnarray}
%
Figure~\ref{fig-g2} shows this second-order coherence function as a function of the laser parameters, and for the case with and without incoherent repumping. Interestingly, the photon statistics can be controlled via the laser field parameters. In particular, for negligible incoherent repumping, the atomic statistics changes from super-Poissonian ($g^{(2)}(0)>1$) to sub-Poissonian ($g^{(2)}(0) <1$) if $\theta$ is varied, as depicted by the short-dashed curve in Fig.~\ref{fig-g2}. Incoherent pumping due to reabsorption of particles, however, restricts the statistics to $g^{(2)}(0) >1$ (see the long-dashed and dotted curves in Fig.~{\ref{fig-g2}}). As it is the case with the trapping state population, also the second-order coherence can be improved, i.e. $g^{(2)}(0) \sim 1$, for smaller samples if $\eta \gg 1$, or for 
$\eta >1$ and larger samples $N \gg 1$.

Finally, we estimate the requirements of the atom beam for collectivity to occur. Present guided ultra-cold atom beams can achieve fluxes of the order of $10^{10}$ atom/s with velocities of order of $1$~m/s, which corresponds to densities of order $n_0=3\times10^{10}$~cm$^{-3}$, or $n_0\lambda^3\approx 0.015$ ($\lambda\approx795$nm in rubidium)~\cite{highflux, pfau}. This is close to the densities at which Dicke superradiance could be observed in an indium sample in a similar level scheme as considered here~\cite{exsf}. There, a $N^2$ dependence of the superradiance intensity was observed starting from $3\times 10^{11}$~cm$^{-3}$, or $n_0\lambda^3\approx 0.027$ ($\lambda\approx450$nm in indium). Thus we conclude that, e.g.,  an increase of the flux or a decrease of the average velocity by about one order of magnitude compared to the results in~\cite{highflux} could allow to enter the regime of collectivity.
Note that the collective decay rates will depend on the sample geometry as $\Gamma^{(col)}_{\alpha} \propto \mu N \Gamma_{\alpha}$ with a geometrical factor  $\mu$~\cite{aei_sup}. Often, $\mu$ can be adjusted to be much smaller than unity. Then, the condition $\Omega \gg \Gamma^{(col)}_{\alpha}$ is satisfied, justifying the secular approximation.

In summary,  we discussed effects of collectivity in a model for loading a magnetic trap. The particle interactions lead to collective decay into a desired trapping state, enhancing the loading performance. We discussed conditions for an efficient increase of the trapping state population, focusing on the ratio of the spontaneous emission rates, the incoherent repumping via photon absorption in the trap, and the pump laser parameters. Finally, we discussed the second order correlation function of the atoms and show that the laser field parameters can lead to a controlled transition between classical and quantum properties of the atoms in the trapping state.

\end{document}